\documentclass[useAMS,usenatbib]{mn2e}

\newcommand{\etal}{\mbox{et\ al.\ }}

\usepackage{rotating}

\title[{\it Swift} captures the prompt emission of GRB\,070616]
{{\it Swift} captures the spectrally evolving prompt emission of GRB\,070616\thanks{This paper is dedicated to the memory of Dr. Francesca Tamburelli who died
during its production. Francesca played a fundamental role within the team
which is in charge of the development of the {\it Swift} XRT data analysis
software at the Italian Space Agency's Science Data Centre in Frascati. She
is sadly missed.}}

\author[Starling \etal 2007]
{R.L.C. Starling$^1$, P.T. O'Brien$^1$, R. Willingale$^1$, K.L. Page$^1$, J.P. Osborne$^1$,\and M.
De Pasquale$^2$, Y.E. Nakagawa$^3$, N.P.M. Kuin$^2$, K. Onda$^4$, J.P. Norris$^{5,6}$,
\and T.N. Ukwatta$^{7,8}$, N. Kodaka$^4$, D.N. Burrows$^9$,
J.A. Kennea$^9$, M.J. Page$^2$, \and M. Perri$^{10}$ and C.B. Markwardt$^{7,11}$\\$^1$Dept. of Physics and Astronomy, University of Leicester, University Road, Leicester LE1
7RH, UK.\\$^2$Mullard Space Science Laboratory, University College London,
Holmbury St. Mary, Dorking, Surrey RH5 6NT, UK.\\$^3$Dept. of
Physics and Mathematics, Aoyama Gakuin
University, 5-10-1 Fuchinobe, Sagamihara, Kanagawa 229-8558,
Japan.\\$^4$Dept. of Physics, Saitama University, Shimo-Okubo 255, Saitama 338-8570, Japan.\\$^5$Denver Research Institute, University of Denver, Denver, CO 80208, U.S.A.\\$^6$Visiting Scholar, Stanford University,
U.S.A.\\$^7$NASA Goddard Space Flight Center, Greenbelt, MD 20771,
U.S.A.\\$^8$Center for Nuclear Studies, Dept. of Physics, The George
Washington University, Washington D.C. 20052, U.S.A.\\$^9$Dept. of Astronomy
and Astrophysics, The Pennsylvania State University, 525 Davey Lab, University
Park, PA 16802, U.S.A.\\$^{10}$ASI Science Data Center, via Galileo Galilei,
00044 Frascati, Italy.\\$^{11}$Dept. of Astronomy, University of Maryland,
College Park, MD 20742, U.S.A.}

\begin{document}
\date{Accepted . Received ; in original form }

\pagerange{\pageref{firstpage}--\pageref{lastpage}} \pubyear{2007}

\maketitle

\label{firstpage}


\begin{abstract}
The origins of Gamma-ray Burst prompt emission are currently not well
understood and in this context long, well-observed events are particularly
important to study. We
present the case of GRB\,070616, analysing the exceptionally long-duration
multipeaked prompt emission, and later afterglow, captured by all the
instruments on-board {\it Swift} and by {\it Suzaku} WAM.
The high energy light curve remained generally flat for several hundred seconds
before going into a steep decline.
Spectral evolution from hard to soft is clearly taking place throughout
the prompt emission, beginning at 285 s after the trigger and extending to 1200 s.
We track the movement of the spectral peak energy, whilst observing a
softening of the low energy spectral slope. The steep decline in flux may be
caused by a combination of this strong spectral evolution and the curvature
effect. We investigate origins for the spectral evolution, ruling out a
superposition of two power laws and considering instead an additional component
dominant during the late prompt emission. We also discuss origins for the
early optical emission and the physics of the afterglow.
The case of GRB\,070616 clearly demonstrates that both broadband coverage and good time
resolution are crucial to pin down the origins of the complex prompt emission in GRBs. 
\end{abstract}

\begin{keywords}
gamma-rays:bursts
\end{keywords}

\section{Introduction}
The prompt emission mechanism for Gamma-ray Bursts (GRBs) is commonly
attributed to internal shocks due to collisions of shells of different
Lorentz factors ejected from the vicinity of a compact object
\citep{Rees}. Later afterglow emission comes from an external shock as the GRB
blastwave decelerates through interaction with the surrounding medium
\citep{Meszaros2}. Though the general picture appears applicable to most GRBs, the details are far from
understood and other models, in particular magnetised flows, have also
been proposed \citep[e.g.][]{Meszaros3,Usov,Kumar1,Zhang1}. 

The advent of the {\it Swift} mission \citep{Gehrels1} has
revealed additional features, for example steep decays and X-ray
flares \citep{Burrows2}, whose properties are consistent with an internal origin
\citep[e.g.][]{Tagliaferri,Nousek,O'Brien,Chincarini,Falcone}. The steeply
decaying phases that directly follow both prompt emission and X-ray flares are
usually interpreted as due to the curvature effect \citep[e.g.][]{Kumar2}
where high latitude emission is delayed with respect to that on-axis. However,
significant spectral evolution is not expected in this model, and new mechanisms must be invoked to explain those
observations that do show spectral evolution during the steep decay phase (Zhang, Liang \& Zhang 2007).

Of particular interest are the longest duration GRBs in which the
relationship between various possible early emission components can be
studied. Observationally, very few GRBs are detected in $\gamma$-rays
for more than 400 seconds as quantified using the T$_{\rm 90}$ parameter. For
example, approximately 0.5\% of the BATSE sample meet this criteria
\citep{Paciesas}. Some of the very long
duration GRBs include those with a pre-cursor or a late flare
\citep[e.g.][]{Price,Nicastro}. Interestingly, the very long GRBs also include
several FRED-like (fast rise, exponential decay) events which can be
spectrally quite soft \citep{Giblin,Zand}.

In the {\it Swift} era it remains true that very few GRBs have a T$_{\rm 90}
>400$ s. The longest is GRB\,060218, which is an unusually long, low
luminosity, spectrally soft GRB \citep{Campana2}. In the first {\it Swift} BAT catalogue \citep{Sakamoto} there are only three others: GRB\,060929, which has a spectrally
soft, late giant flare \citep{Palmer}; GRB\,070129, for which
BAT triggered on a precursor \citep{Krimm}; and GRB\,070616.

Here we present data for the case of GRB\,070616, in which the prompt emission
shows a very
complex multipeaked structure, leading to one of the longest prompt
emission durations ever recorded. We take advantage of extensive coverage of such a long
burst by the {\it Swift} Burst Alert Telescope \citep[BAT;][]{Barthelmy}, the X-Ray Telescope \citep[XRT;][]{Burrows1} and the
UV/Optical Telescope \citep[UVOT;][]{Roming}. Combining data from
{\it Swift} and {\it Suzaku} \citep{Mitsuda} we study the evolution of the prompt
emission spectrum, following the temporal variability of the peak
energy, and study the development of the afterglow component.

In Section 2 we describe the {\it Swift} and {\it Suzaku}
observations. In Section 3 we outline the X- and
$\gamma$-ray temporal characteristics. In Section 4 we
model the X-ray to $\gamma$-ray spectra and investigate spectral
evolution. We discuss the discovery of an optical transient in the
UVOT images in Section 5 and go on to model the spectral energy
distributions in Section 6. In Section 7 we present possible
interpretations for our findings, and discuss them in the context of
the blastwave model and in comparison with other GRBs, with summary and
conclusions in Section 8.

\section{Observations}
On 2007 June 16 at 16:29:33 UT (hereafter T$_{\rm 0}$), the {\it Swift} BAT triggered on and
located GRB\,070616 (trigger=282445, Starling et al. 2007a). This was an image
trigger, where the GRB is detected by forming an image from the collected
counts which is then searched for a new point source.
{\it Swift} slewed immediately to the BAT location.
T$_{90}$
(15--350 keV) is $402\pm10$ s (estimated error including systematics), which
is among the longest T$_{90}$ values for BAT GRBs.

The XRT began observing the field at 16:31:44 UT, 131 seconds after the
BAT trigger; a bright, fading and uncatalogued X-ray source was clearly
detected. 
XRT observed the source for a total exposure of 144.5 ks out to
T$_{0}$+3.7$\times$10$^5$ s, beginning in Windowed Timing (WT) mode at
T$_{0}$+137 s and
continuing in Photon Counting (PC) mode at T$_{0}$+976 s.
Using 899 s of overlapping XRT PC mode data and UVOT $V$ band data, we
obtain an astrometrically corrected X-ray position \citep[using the USNO-B1 
catalogue,][]{Goad2} of
RA (J2000) = 02h 08m 36.59s;
Dec (J2000) = +56$^{\circ}$ 56$'$ 43.8$''$,
with an error of radius 2.4 arcseconds (90\% containment).
The error circle contains the bright star USNO-B1.0 1469-0076513.

The UVOT took a finding chart exposure of 100 seconds with the White
filter starting at T$_{\rm 0}$+142 s, and continued with a long $V$ band
exposure followed by cycling through all seven filters.
The count rates in $V$ and $U$ band exposures between 250 and 1050
seconds after the trigger show a significant excess (at $\geq$3$\sigma$) with respect to late observations
($>$1 day after the trigger) at the position of the USNO-B1.0 star. We
discuss this further in Section 5. 

The {\it Suzaku} Wide-band All-sky Monitor \citep[WAM;][]{Yamaoka}, one
element of the Hard X-ray Detector \citep[HXD;][]{Takahashi} with total bandpass 50 keV to 5 MeV, detected GRB\,070616 at 16:31:50 UT,
137 s after the {\it Swift} BAT trigger. The observations cover 140 s of the GRB emission, truncated by
passage into the South Atlantic Anomaly. With these data alone we put a limit
on the GRB duration of T$_{90}$ of $>$ 112 s over 150--700 keV, consistent with the BAT (15--350
keV) T$_{90}$. 
\begin{figure}
\begin{center}
\includegraphics[width=8.8cm, angle=0]{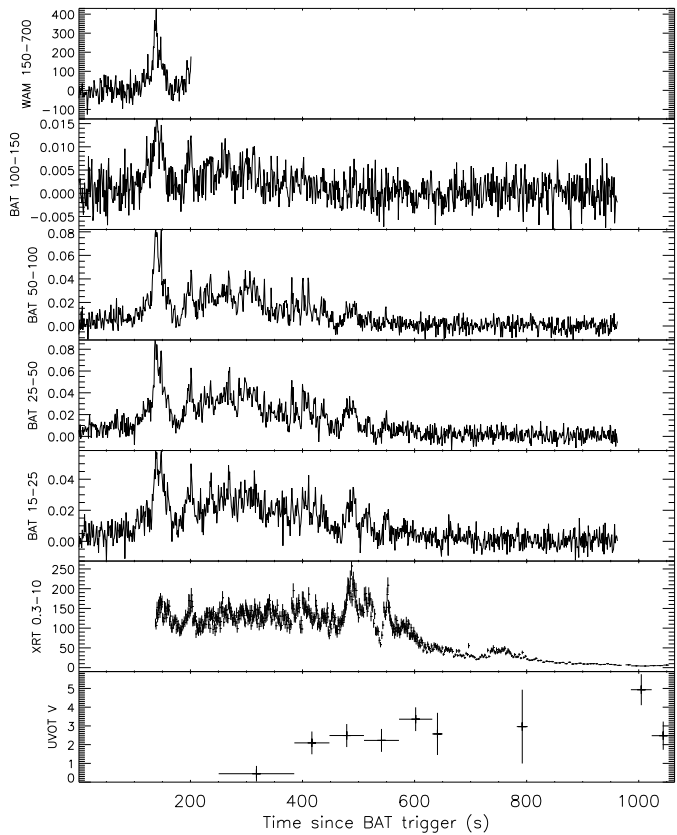}
\caption{Light curves for the {\it Suzaku} WAM and all {\it Swift} instruments
  are shown in count s$^{-1}$. The
  instrument and, where appropriate, the energy range used in keV, are given in the
  y-axis labels. We show here the WAM 1 s light curve and the mask-weighted 1 s light curves for the BAT in 4
  individual energy ranges. The UVOT observed in all seven of its filters, but
we only show the $V$ band here (500--560 nm), in which the majority of the
detections were obtained, for clarity; USNO-B1.0 1469-0076513 has been subtracted.}
\label{all lightcurves}
\end{center}
\end{figure}

The {\it Swift} data were processed with the standard {\it Swift} data reduction
pipelines, and spectra and light curves were extracted with xselect. All XRT
spectra are grouped such that a minimum of 20 counts lie in each bin, and X-ray
light curves have a minimum of 15 source region counts per bin. We used a source extraction
region of box width 40 pixels for XRT spectra in WT mode (1 pixel = 2.36$''$). In the first orbit of PC mode
data, emission from the source was piled up, and spectra were extracted using
a 20 pixel radius annulus
excluding the inner 4 pixels in radius; a correction for this was made to the
ancilliary response file. Thereafter a circular source region of radius 17 pixels
was employed. Background spectra were extracted from a nearby source-free
region of radius 86 pixels. 
Spectral fitting was
performed in Xspec \citep{Arnaud} using version 008 response files.
Errors are given at the 90\% confidence level throughout, unless otherwise stated.

\section{Temporal characteristics of the prompt emission}
\begin{figure}
\begin{center}
\includegraphics[width=6cm, angle=-90]{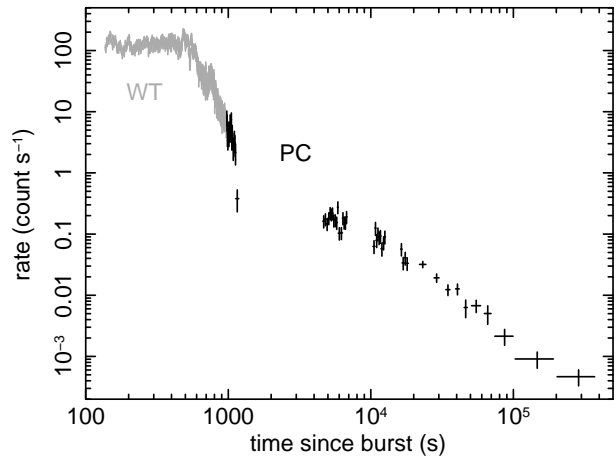}
\caption{The XRT 0.3-10 keV count rate light curve, plotted in log space.}
\label{XRT lightcurve}
\end{center}
\end{figure}

Light curves for the prompt emission in all observed bands are shown in
Fig. \ref{all lightcurves}. The $\gamma$-ray light curves show no strongly
peaked or variable emission 
around T$_{\rm 0}$, as is common for BAT image triggers such as this. BAT
triggered instead on a gradual rise which lasted approximately 100 s before the
first and strongest peak, centred at T$_{\rm 0}$+120 s. Thereafter multiple blended peaks continue
the prompt emission out to T$_{\rm 0}$+500--600 s. At this point the $\gamma$-ray
emission appears to return to the count rate at which it began at T$_{\rm 0}$, whilst the X-ray
emission begins a steep decay lasting until T$_{\rm 0}$+1200 s (Fig. \ref{XRT lightcurve}), and the $V$ band optical emission continues a
steady rise (Fig. \ref{all lightcurves}).

The X-ray observations cover much of the time period over which BAT could
detect the source, and we constructed a joint BAT-XRT light curve to compare emission from
the two bands. 
This was done by performing joint power law fits
to fine time-sliced BAT-XRT (WT) spectra, and
extrapolating the BAT spectra to the XRT energy band. The main peaks in the
multipeaked prompt emission are found to be
temporally coincident, strongly suggesting that the X-ray and $\gamma$-ray
emission come from the same component: the GRB prompt emission. In contrast,
the $V$ band optical data show a slowly rising light curve from T$_{\rm 0}$+250--1000 s which is not mimicked in
any of the high energy bands (Fig. \ref{all lightcurves}).

We performed a lag analysis \citep[e.g.][]{Norris}
over two intervals in the BAT light curve: $\sim$70 s covering the largest
peak, and $\sim$155 s covering the flat multipeaked structure following
the largest peak. We use the four
BAT channels 15--25 keV (channel 1), 25--50 keV (2), 50--100 keV (3) and
100--350 keV (4) and 256 ms time binning. For the first interval we find the following lags: \\
3$\rightarrow$1 = 220$^{+90}_{-150}$ ms ; 4$\rightarrow$2 = 410$\pm$200 ms \\
and in the second interval the lags have decreased to:\\
3$\rightarrow$1 = 64$^{+64}_{-72}$ ms ; 4$\rightarrow$2 = 0$^{+80}_{-112}$ ms.\\
The time resolution of the analysis prevents accurate measurements of lags $<$256 ms,
causing the relatively large errors reported here.

We attempted to model the underlying temporal
behaviour of the prompt X-ray emission by fitting a multiply broken power law model to the XRT light curve, as is often done for complex GRB light
curves. We find the following temporal slopes, $\alpha$ ($F \propto t^{-\alpha}$), and break times
$T_{\rm bk}$: $\alpha_1$ = -(0.08$^{+0.03}_{-0.01}$) up to $T_{\rm bk,1}$
= 524$^{+2}_{-3}$, $\alpha_2$ = 4.87$^{+0.11}_{-0.08}$ up to $T_{\rm bk,2}$ =
999$^{+18}_{-24}$ and $\alpha_3>10$ up to 1200 s (the steepest
segment of the decay has a slope so large when fit with a
simple power law it is difficult to accurately measure). Between the steep
slope and the first probable afterglow emission, T$_{\rm 0}$+1130 to T$_{\rm
  0}$+5800 s, we measure $\alpha_{\rm plateau}$ = 0.5$^{+0.3}_{-0.2}$ using a
power law fit to the last point on the steep decay and the second orbit of PC mode data.
The afterglow
temporal slope from $>$4600 s using this method is $\alpha_4$ = 1.46$\pm0.05$.
\begin{figure}
\begin{center}
\includegraphics[width=6cm, angle=-90]{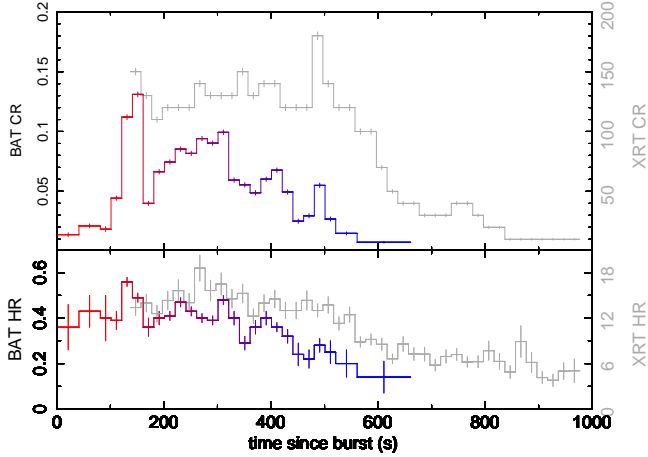}
\caption{Count
  rate (CR, upper panel) and hardness ratio (HR, lower panel) against time for XRT (1-10 keV/0.3-1 keV, 10 s bins,
  grey) and BAT (15-50 keV/50-100 keV, 10, 20 or 50 s bins, colour scale).}
\label{HR}
\end{center}
\end{figure}
\begin{figure}
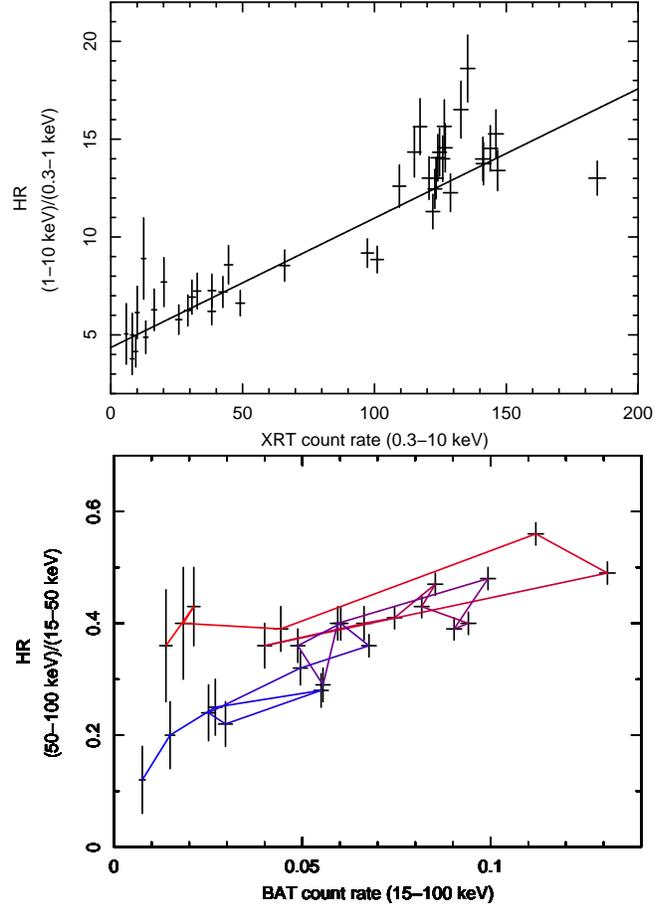

\begin{center}
\includegraphics[width=6cm, angle=-90]{fig4a.ps}
\includegraphics[width=6cm, angle=-90]{fig4b.ps}
\caption{Hardness ratio against
  count rate for XRT (upper) and BAT (lower). The evolution of
  both quantities with time is shown by the colour gradient in the BAT panel, going
  from red to blue as time increases from T$_{\rm 0}$ to T$_{\rm 0}$+650
  s. This colour gradient is identical to that shown in Fig. \ref{HR}. The general
  trend of harder spectra at higher count rates is visible, and within this we
  see individual loops of soft-hard-soft evolution. Each loop shows
  the effect of the larger individual flares.}
\label{HRvsCR}
\end{center}
\end{figure}

For the XRT and BAT we obtained hardness ratios as a
function of time (Fig. \ref{HR}). 
The BAT hardness ratio (15--50 keV/50--100 keV) remained approximately constant
until T$_{\rm 0}$+285 s when the spectra softened
significantly over the remainder of the $\gamma$-ray observations. The XRT hardness
ratio (1--10 keV/0.3--1 keV) shows the same behaviour, with the spectral
evolution beginning at approximately T$_{\rm 0}$+500 s (see also Fig
\ref{curvatureplus} in Section 7.1). In both bands the emission begins with a hard
spectrum, evolving to become softer. The source is spectrally harder at higher
count rates, as shown in Fig. \ref{HRvsCR}. 
In the lower panel of Fig. \ref{HRvsCR} we show
that flares themselves are not driving this correlation, although they are
consistent with it, but that there is an
overall trend in the hardness ratios with only the smaller scale changes corresponding
to hardening within individual
flares (individual flares can be seen as loops, following the colour scale
from red through to blue which represents the passage of time).

\begin{table}
\caption{~X-ray spectral fits, using the XRT energy range 0.3--10 keV, and over varying time
  ranges covering the prompt phase: T$_{\rm 0}$+137--1200 s. We test the single
  power law model with various absorption columns
  applied. Fstat refers to the F-test statistic
  when comparing the fits with absorbing column allowed to vary against those
  fixed at the Galactic column of 0.35$\times$10$^{22}$ cm$^{-2}$.}
\begin{center}
\begin{tabular}{c|cccc}
Time & $N_{\rm H}$ & $\Gamma$ & $\chi^2/dof$
&F-stat \\ \hline
s & $\times$10$^{22}$ cm$^{-2}$&  & &  \\ \hline \hline
137-963&0.35 & 1.26$\pm$0.01 &  897/720 &\\ \hline
137-237&0.35 & 1.06$\pm$0.03 &  346/387 &\\
237-337&0.35 & 0.99$^{+0.02}_{-0.03}$ &398/409& \\
337-437&0.35 & 1.13$^{+0.02}_{-0.03}$ &393/410 &\\
437-537&0.35 & 1.30$\pm$0.02  & 489/396&\\
537-637&0.35 & 1.59$\pm$0.03  & 288/284 &\\
637-737&0.35 & 1.84$\pm$0.05 &  150/127 &\\
737-837&0.35& 1.85$^{+0.05}_{-0.06}$ & 109/111&\\
837-963&0.35 & 1.97$\pm$0.10  & 59/51 &\\
976-1200&0.35& 2.2$\pm$0.3 &  2.44/6 &\\\hline
137-963&0.40$\pm$0.01&1.34$^{+0.01}_{-0.02}$ & 824/719&6$\times$10$^{-15}$
\\\hline
137-237&0.39$\pm$0.03&1.09$^{+0.06}_{-0.03}$ 
&341/386 &0.02\\
237-337&
0.42$^{+0.03}_{-0.07}$&1.07$^{+0.03}_{-0.05}$ & 385/408&2$\times$10$^{-4}$\\
337-437& 0.42$\pm$0.03&1.21$\pm$0.04 & 375/409&1$\times$10$^{-5}$\\
437-537& 0.49$\pm$0.03 & 1.50$^{+0.03}_{-0.05}$ &397/395&1$\times$10$^{-19}$\\
537-637& 0.42$^{+0.04}_{-0.02}$ &
1.71$^{+0.05}_{-0.04}$  &  258/283&2$\times$10$^{-8}$\\
637-737 & 0.42$^{+0.05}_{-0.04}$ & 1.96$\pm$0.09&  142/126 &0.01\\
737-837& 0.39$^{+0.06}_{-0.03}$&1.91$^{+0.09}_{-0.06}$& 104/110 &0.02\\
837-963 & 0.30$^{+0.05}_{-0.07}$ & 1.86$^{+0.11}_{-0.16}$  & 56/50&0.11\\
976-1200& $<$0.9 & 2.1$^{+0.7}_{-0.5}$ & 2.35/5 &0.68\\ \hline
137-237&0.40& 1.12$\pm$0.03 & 341/387 &\\
237-337&0.40& 1.05$^{+0.02}_{-0.03}$ &  385/409& \\
337-437&0.40& 1.19$^{+0.02}_{-0.03}$ & 376/410& \\
437-537&0.40& 1.37$^{+0.03}_{-0.02}$& 431/396& \\
537-637&0.40& 1.68$^{+0.03}_{-0.04}$& 261/284&\\
637-737&0.40& 1.93$^{+0.07}_{-0.06}$ &  143/127& \\
737-837&0.40& 1.95$^{+0.05}_{-0.06}$ &  104/111& \\
837-963&0.40& 2.06$\pm$0.10 & 64/51& \\
976-1200&0.40& 2.3$\pm$0.3 &  2.55/6& \\\hline
\end{tabular}
\label{tab:XRTspecfits2}
\end{center}
\end{table}

\section{Spectral fitting}
\subsection{X-ray}
To investigate the spectral evolution seen in Figs. \ref{HR} and \ref{HRvsCR}, we began
by characterising the XRT X-ray spectrum for the prompt emission
interval. All results are given in Table \ref{tab:XRTspecfits2}. To test for the
presence of intrinsic absorption, we extracted a time-averaged spectrum over the time
range T$_0$+137--963 s. We fit the spectrum with a
power law model with absorption set initially to the Galactic
value of $N_{\rm H,Gal} = 0.35 \times 10^{22}$ cm$^{-2}$ \citep[LAB survey,][]{Kalberla} and then allowed to go free. The inclusion of a small column of intrinsic
absorption, amounting to $5 \times 10^{20}$ cm$^{-2}$ at $z=0$, is required according to
the F-test (F-statistic = 6$\times$10$^{-15}$, Table \ref{tab:XRTspecfits2}).

We then extracted X-ray spectra in 100 s time bins from T$_{\rm 0}$+137 to T$_{\rm 0}$+837 s, and two further longer time bins of sizes 126 s and 224 s to extend
coverage to the end of the steeply decaying phase (T$_{\rm 0}$+1200 s).
We fit these spectra with a power law plus Galactic+intrinsic absorption
(amounting to a total column of $N_{\rm H} = 0.4 \times 10^{22}$ cm$^{-2}$). 
We also allowed the intrinsic absorption column
to vary to test for any temporal changes which may be caused by e.g. ionisation by the
GRB emission, but found no variation confirming that the time-averaged $N_{\rm H,
intrinsic}$ that we have applied is appropriate throughout.
We observed the power law slope to change from
$\Gamma\sim1.1$ to $\Gamma\sim2.3$ in $\sim$800 s, indicating that a spectral
softening occurred in the X-ray band. This begins after the T$_{\rm 0}$+337--437 s
segment, and corresponds to the time of the change in the X-ray hardness ratio
(see Fig. \ref{HR}).

We also fit a broken power law model to the same data, initially with Galactic
absorption only.
The broken power law model provides statistically equal or better fits than the
absorbed power law
model in most cases, but this
may be a symptom of the need for intrinsic absorption, and/or possibly
additional spectral curvature.
Adding $5 \times 10^{20}$ cm$^{-2}$ of intrinsic X-ray absorption does not
significantly improve the majority of the broken power law fits. Although it is
difficult to statistically distinguish between a power law plus intrinsic extinction and a
broken power law, we find that the break energies and power law slopes in
the latter model move seemingly randomly in time whereas the steepening in the
single power law model matches the hardness ratio behaviour very well. In
general, we expect that there will be some material within the host galaxy
that lies in front of the GRB and along our line-of-sight causing detectable
X-ray absorption \citep[e.g.][]{Campana1}. We therefore do not consider the broken
power law model further in fits to the XRT data.

The X-ray afterglow during the second orbit of PC mode data (T$_0$+4608--6976 s) shows a soft spectrum which is well fitted by a single
absorbed (Galactic+intrinsic as above) power law with
$\Gamma=2.53^{+0.20}_{-0.19}$ ($\chi^2$/dof = 11/13).
No spectral variation is seen through the afterglow phase when comparing the
above time range with data from T$_0$+10~ks onwards which are well fit with a power law
of photon index $\Gamma=2.62^{+0.19}_{-0.17}$ ($\chi^2$/dof =14/16).

\subsection{ X-ray to $\gamma$-ray}
We now extend the above analysis to include the higher energy BAT spectrum.
Firstly we extracted a time-averaged BAT-XRT spectrum covering the prompt emission phase from
the start of XRT observations to the end of source visibility with the BAT
(137--963 s). We fit this with three models: a power law, broken
power law and the Band function \citep{Band}. The normalisation for each
instrument is always tied such that they are equal. In the Band function fits
we fixed the high energy power law slope to $\beta=2.36$ as found in the
{\it Suzaku} spectrum \citep[][and adopting $F_{\nu} \propto \nu^{-\beta}$]{Morigami}, since the energy range with slope $\beta$ is expected, at
least initially, to lie above the BAT band. Each model was also absorbed by
either the Galactic column alone, or the
previously determined Galactic+intrinsic column. Again inclusion of the small
amount of intrinsic absorption improved the power law and Band fits, but was
not a significant improvement for the broken power law fits. 
Results are given in Table 2.
A single power law is a poor representation of the time-averaged joint spectrum, and we
find the Band function or a broken power law provides a better fit.

\begin{figure}
\begin{center}
\includegraphics[width=6cm, angle=90]{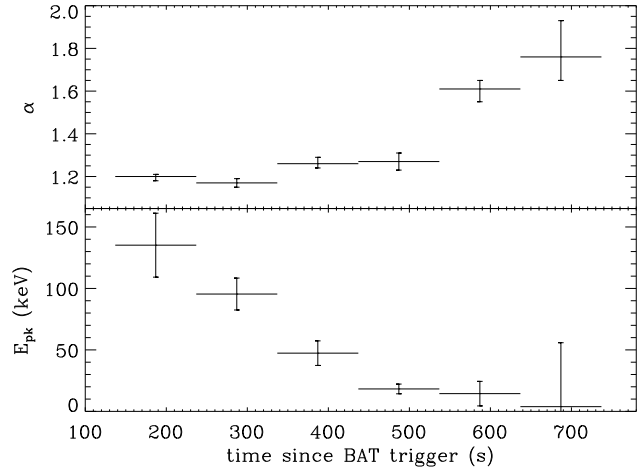}
\caption{The Band model fitted to the BAT-XRT spectra in seven 100 s
  intervals, where the fitted parameters $\alpha$ (upper panel), and $E_{\rm 0}$ combine
  to give $E_{\rm pk}$ (lower panel). 90\% errors on $E_{\rm pk}$
  have been calculated using average symmetrical error bars for $\alpha$ and $E_{\rm 0}$.}
\label{Epkband}
\end{center}
\end{figure}

To follow the spectral evolution, we time-sliced the BAT data
into six
100 s bins covering T$_{\rm 0}$+137--737 s, corresponding to some of those extracted for the XRT.
We fit the three
models listed above to the joint BAT-XRT spectra for each time bin, this time
absorbed by the combination of Galactic and intrinsic absorption. 
The single power law model shows the same softening of the
spectral slope with time as was seen for the time-sliced XRT data.
The broken power law or the Band function again provide a better fit than a
single power law model. 
The broken power law model fits suggest a fairly constant break in the middle of the
X-ray band at $\sim$4--5 keV (we note that the fits are biased towards the X-ray
band where there are the greatest number of counts). 
In the Band function fits the peak
energy is derived from the free parameters $\alpha$ (low energy power law
index) and $E_{\rm 0}$ (characteristic energy) and the fixed
parameter $\beta = 2.36$ (high energy power law slope) using 
\begin{equation}
E_{\rm pk} = E_{\rm 0} (2-\alpha)
\end{equation}
(see Band et al. 1993 for a detailed description of the Band function; errors on $E_{\rm pk}$ are calculated using average symmetrical error bars). $E_{\rm pk}$
can be well constrained and is observed to move to lower
energies with time from 135 keV down to 4 keV in $\sim$600 s
(Fig. \ref{Epkband}, Table 2),
while the spectral slope
$\alpha$ also varies gradually, softening with time.

\begin{figure}
\begin{center}
\includegraphics[width=6cm, angle=-90]{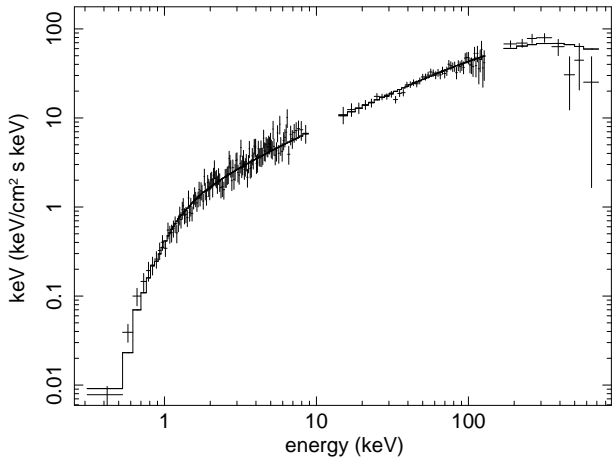}
\caption{Spectral energy distribution of the WAM, BAT and XRT spectra at
  T$_{\rm 0}$+133--159 s, with best fitting absorbed Band function. The peak of the spectrum
  can be seen close to the boundary between the BAT and WAM bandpasses.}
\label{WAMBATXRTBand}
\end{center}
\end{figure}

To confirm the validity of the high energy spectral slope we fixed in the
previous Band function fits and the position of the peak energy derived from fits to the {\it Swift} data alone, we performed power law fits to the {\it Suzaku}
WAM spectrum. 
The constraints we derive on the power law slopes are consistent with the value
$\beta=2.36$ used in previous analysis (see Table 2).
We then performed joint fits for WAM, BAT and XRT. 
The results allow us to confirm the previous findings using a broader-band
spectrum. In particular, previous BAT-XRT fits suggest the peak energy of the spectrum lies in the
{\it Suzaku} energy band at these times ($\le$250 s) and is therefore more
accurately measured in the WAM-BAT-XRT joint fit. Results are reported in
Table 2.
The absorbed Band function fit to the broadband data is shown in the $E^2 F_{E}$ plot (energy equivalent of $\nu F_{\nu}$) in Fig. \ref{WAMBATXRTBand}. 

\section{Identifying the optical afterglow}
\begin{figure}
\begin{center}
\includegraphics[width=8cm, angle=0]{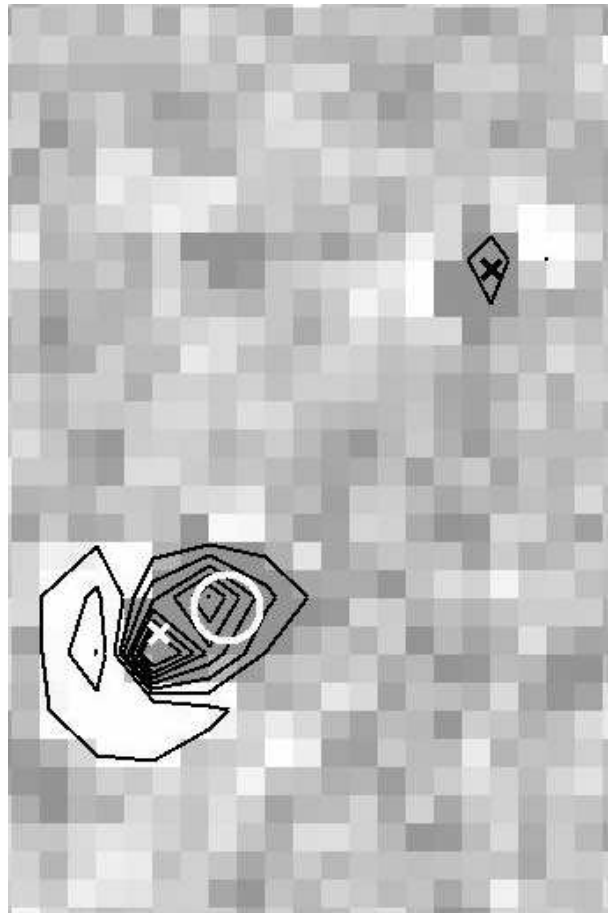}
\caption{UVOT V band
    image subtraction revealing the fading optical counterpart (negative areas are white, while positive
 areas are shades of grey). The white
    circle is the OT position and the white `x' marks the centre of the star
    USNO-B1.0 1469-0076513, as
    measured from both $V$ and $U$ bands. Contours show both
    the residuals of the nearby star and the OT.
The black `x' indicates the
    position of nearby field star USNO-B1.0 1469.0076499. 1 pixel = 1$''$.
    }
\label{subtraction}
\end{center}
\end{figure}

An optical transient (OT) was not immediately identified in the {\it Swift} UVOT
images owing to
superposition with the bright ($V$ = 14.4) star USNO-B1.0 1469-0076513, whose
colours are consistent with those of a hot star. 
The blended image of the star plus the OT is extended
compared to other stars in the field. 
In both the UVOT $V$ and $U$ bands the OT was separated
from the nearby star by subtracting a properly scaled later
image from earlier images.
For the $V$ band subtraction we used the sum of the first and third exposures
(T$_0$+250--650 s and T$_0$+986--1152 s) and a sum of 2 later exposures covering T$_0$+178.2--185.1 ks. The $U$ band subtraction used a sum of the first 2
exposures (T$_0$+704--723 s and T$_0$+866--874 s)
and a sum of 3 later
exposures covering
T$_0$+97.2--98.1 ks. The scaling was based on the
total exposure times and all images were mod-8 and aspect-corrected, and rebinned to 2$\times$2 UVOT
sub-pixels (1$''$$\times$1$''$).
The subtracted images, of which the $V$ band is shown in Fig. \ref{subtraction}, reveal
the sources which have varied between the considered epochs.
Due to minor changes in the average point spread function (PSF) of the stellar
source during the orbit, count rate noise in the original
images, and the combined effect of uncompensated
coincidence-loss where the OT and stellar PSFs overlap, the
stellar image could not completely be removed and is present
as a low intensity peak and ring of opposite sign. Some residual
of other bright stars can thus also be found in the subtraction.
The net images in the
$V$ and $U$ filters differ slightly due to different PSF and
coincidence-loss effects in each band. These complications have only a minor effect on the
determination of the OT position, which is found from a combination of the
subtracted images of both bands.
We find a best position for the
afterglow at RA (J2000) = 02h 08m 36.37s;
Dec (J2000) = 56$^{\circ}$ 56$'$ 44.1$''$,
with an uncertainty of 1.2$''$ (radius, 90\% confidence).
Using a 4$''$ radius extraction region
and a nearby source-free background region, we determine the count rates and flux densities for
the OT after subtracting those of the star (Table \ref{tab:uvot}). 
The OT is detected in the $U$ and $V$ bands at $>$3$\sigma$, and at lower
significance in the $B$ band. It is not detected in the UV filters and has disappeared below detection limits in exposures
taken after 1147 s in $V$, 739 s in $B$ and 870 s in $U$. We do not use the white filter here due to high
coincidence losses. Detection of this source in the $U$ band indicates a
redshift less than $\sim$3.

A limit on the near-infrared magnitude was found using the
1.34m ground-based TLS Tautenburg telescope using ISIS image
subtraction \citep{Alard} of two epochs, finding a conservative limit on the
magnitude of $I>19$ at 0.3 days
after the trigger \citep[K. Wiersema \&
D.A. Kann, private communication; see also][]{Kann}.

To support the argument that the OT is the optical counterpart to the GRB, we performed
an identical analysis on another star in the same field of view. This star reveals no
change of flux throughout the observations, ruling out
instrumental effects as the cause of the flux variations seen in the GRB OT. Observations carried out for a further 2 days
suggest the star superposed on the OT position can be considered a constant source, 
making it highly unlikely that
the star in the XRT error circle flared at such
time that it mimicked a GRB afterglow. 
\begin{table}
\addtocounter{table}{+1}
\caption{UVOT background subtracted measurements and limits for the optical afterglow candidate, performed with
  a 4 arcsec aperture and with the contribution from the USNO-B1.0 star removed. All errors are 1$\sigma$ and upper limits
  3$\sigma$. Galactic extinction of
  $E(B-V)\sim0.4$ \citep{Schlegel} has not been accounted for.}
\begin{center}
\begin{tabular}{ccccc}
 Band &    T$_{\rm mid}$        &   T$_{\rm exp}$ & Count/s &   Flux ($\mu$Jy) \\    
\hline \hline
 {\it V} &  317.8 & 135 &0.45$\pm$0.42 &113$\pm$105\\   
&      416.7   &  62.6  &2.09$\pm$0.61&  530$\pm$154\\
&      478.9   &  61.5 &2.49$\pm$0.61 & 630$\pm$155 \\
&      540.9  &   62.4  &2.23$\pm$0.61&  564$\pm$155\\
&      601.9 &    59.3 &3.36$\pm$0.63 & 850$\pm$160\\
&      640.8  &   18.4  &2.57$\pm$1.12&   651$\pm$284\\
&      792.2    &  19.5  &2.96$\pm$1.97 & 749$\pm$277 \\         
 &     1005 &     37.1   &4.93$\pm$0.83&  1248$\pm$209\\
 &    1044 &    41.0   &2.48$\pm$0.75 &627$\pm$191\\
 &   1085  &    41.1 & 2.45$\pm$0.75  & 620$\pm$190\\
 &  1126 &      41.0  &2.48$\pm$0.75& 627$\pm$191\\
&5942     &200&$<$0.99 &$<$253 \\
&11729 & 885 &$<$0.49 & $<$124   \\
\hline
{\it B}&734.0 & 10.0 &3.81$\pm$2.13 &309$\pm$173 \\
 &6044& 393&$<$1.04& $<$84   \\
 &18286 &  615&$<$0.83&        $<$67.5   \\
\hline
{\it U}&714.0  &9.8&2.74$\pm$1.30& 168$\pm$80\\
&865.0 &9.8 &2.48$\pm$1.29 &152$\pm$79\\
&5839	 &393      &$<$0.86&$<$53   \\
&17516    &885       &$<$0.56&$<$34    \\
\hline
{\it UVW1} &765.0&39.8 &$<$1.47&$<$143  \\
\hline
{\it UVM2} & 740.0&39.8 &$<$0.84& $<$104\\
\hline
{\it UVW2} &767.5 &19.0&$<$1.68& $<$139 \\
\hline
\end{tabular}
\label{tab:uvot}
\end{center}
\end{table}
\section{Spectral energy distributions}
We constructed three optical to X-ray spectral energy distributions (SEDs)
spanning the prompt emission phase and one SED in the afterglow phase using
the ISIS spectral fitting package \citep{Houck} and fitting in count space
using the method described in \cite{Starling2}. Throughout these fits we fixed
the total X-ray absorption (Galactic+intrinsic) to the value of $N_{\rm
  H}=0.4\times10^{22}$ cm$^{-2}$ derived from the X-ray spectral fits. Galactic
extinction is also included in all fits with $E(B-V)=0.4$ \citep{Schlegel}. We adopt
the Small Magellanic Cloud (SMC) extinction curve for fits to the intrinsic optical
extinction in the host galaxy\footnote{The SMC extinction curve is generally a better fit
 to GRB host galaxy extinction than that of the Milky Way or the Large
 Magellanic Cloud \citep[e.g.][]{Starling2,Schady}. The SMC has the lowest metallicity of these three
  nearby galaxies for which extinction curves have been derived, more closely
  resembling the low metallicity GRB host galaxies. We also note that Solar
  metallicity is assumed for the X-ray absorption measurements which means
  the adopted $N_{\rm H, intrinsic}$ value is likely a lower limit.}, allowing for distances to
the source of $z=$ 0--3 (see Section 5).

The afterglow spectral energy distribution at 5670 s comprises UVOT $B$ and $V$
upper limits and XRT data. Fits
to the SED are consistent with the synchrotron
model, if there is a break in between the optical and X-ray bands (discussed
further in Section 7.3). Fixing the difference
between the power law slopes in this broken power law model to 0.5, as expected
for a cooling break, implies that some intrinsic absorption is also
needed. Fitting to the optical upper limits and X-ray data we can set a lower limit
to the required optical extinction of 
$E(B-V) \ge 0.1$. This amount of intrinsic optical extinction lies at the high
end of the observed distribution \citep[e.g.][]{Starling2}, whilst the intrinsic X-ray absorption is relatively
small for a GRB \citep[e.g.][]{Campana1}.

The optical emission observed at earlier times is not necessarily associated with the
high energy prompt emission. In creating the SEDs for these prompt phases we
assume the optical and X-ray/$\gamma$-ray emission do come from the same component, and determine what spectral shape
and amount of intrinsic extinction would be required in this scenario. 
The first of the prompt phase SEDs at 284 s includes UVOT $V$ band, XRT and BAT data. We fit a
Band function to the spectrum with the high energy photon index $\beta$ fixed
to 2.36. The
availability of optical and high energy data provides a rare opportunity for Band function fits to a broadband
SED. The $V$ band datapoint lies below the best possible fit
to the whole SED. To obtain an acceptable
fit either intrinsic extinction would have to be present at the level of
$E(B-V)\ge0.05$ or both extinction and a further spectral break could be present. 

\begin{figure}
\begin{center}
\includegraphics[width=6cm, angle=-90]{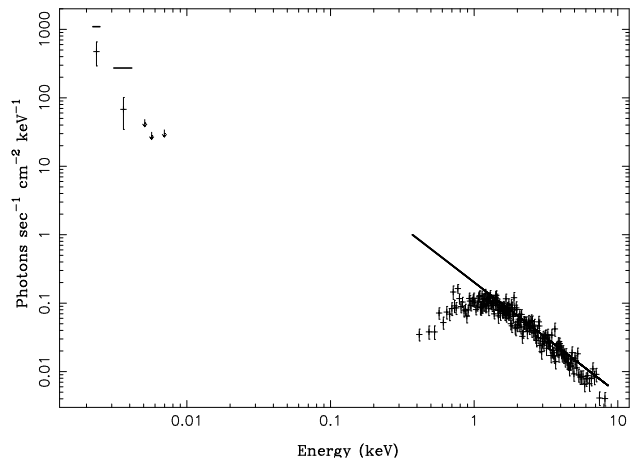}
\caption{The SED at 770 s after the trigger: XRT and $V$ and $U$ band data and UV filter
  upper limits. The unfolded (unabsorbed) model
  shown as a solid line is a power law fit to the observed $V$, $U$ and X-ray
  data, showing the overprediction of the low energy spectrum.}
\label{allseds}
\end{center}
\end{figure}

The second epoch SED at 770 s includes UVOT $U$ and $V$ band and XRT data. We
fit an absorbed
power law to the spectrum. The power law model overpredicts the optical data, illustrated in
Fig. \ref{allseds}, and allowing for some intrinsic extinction decreases the
$\chi^2_{\rm reduced}$ value by $\sim$50\%.
In order to obtain an acceptable
fit either intrinsic extinction would have to be present at the level of
$E(B-V)$ = 0.03--1.40, or we have a spectral break plus extinction. For the latter
model we find a best
fitting break energy of 0.7$\pm$0.2 keV, $\Gamma_1$ = 1.4$^{+0.3}_{-0.2}$, $\Gamma_2 =
$ 1.97$\pm$0.05 consistent with BAT-XRT Band function fits (Section 4.2) and
$E(B-V) \le$ 0.5. The power law slopes differ by $\sim$0.5 consistent with a
cooling break inferring that the cooling break moves to lower energies with
time. Both single and broken power law fits are, however, statistically acceptable and indistinguishable.

The third epoch SED at 1082 s is the least constrained: this SED covers the
final part of the steep X-ray decay phase of the light curve where the X-ray
source has become fainter and XRT was operating in PC mode. A joint fit of
the UVOT $V$ band data and the X-ray spectrum
manages to accommodate the single optical point without inclusion of any
intrinsic extinction, with a slope of $\Gamma=1.76^{+0.05}_{-0.07}$ and $\chi^2$/dof = 8.6/7, but is inconsistent with the fit to
the X-ray data alone of $\Gamma=2.3\pm0.3$ (Table \ref{tab:XRTspecfits2},
Section 4.1) and with the afterglow SED fits. Including an intrinsic
extinction component the power law photon index is then
best fit with $\Gamma=2.3\pm0.3$ and the range of allowed extinction values is
$E(B-V)=$ 0.5--2.2. More complex models cannot be tested with this dataset.

\section{Discussion}
\begin{figure}
\begin{center}
\includegraphics[width=6cm, angle=-90]{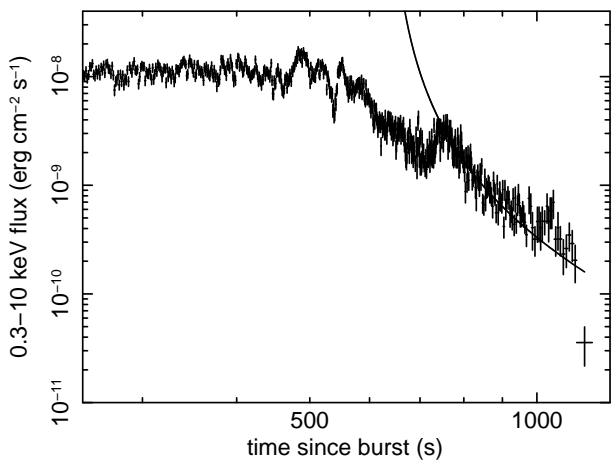}
\caption{The XRT light curve and fit to the energy injection time for the last
  occurring X-ray flare using the curvature model. A variable count rate to flux
  conversion was applied. This fit gives a T$_{\rm 0,flare}$ of 632$^{+11}_{-12}$ s.}
\label{Tzero}
\end{center}
\end{figure}
In GRB\,070616 we have the rare opportunity to track the detailed
spectral and temporal evolution from $\gamma$-rays to X-rays throughout
the entire prompt emission phase. Hence we devote the majority of this
discussion to the possible origins of the complex prompt emission, as
well as the early optical emission, using the observational results
given above. We also provide a brief discussion of the afterglow
characteristics in the framework of the standard synchrotron model.

\subsection{The prompt emission mechanisms}
The prompt emission of GRB\,070616 is atypical of GRBs in that the
emission rises relatively slowly over about 100 seconds to a peak,
then persists at a fairly constant level for 150 s in $\gamma$-rays and
350 s in X-rays before
showing a rapid decline. Throughout the $\sim$ constant phase the light
curve resembles a large number of flares or flickering superimposed on
an underlying constant intensity emission. During this prompt phase we observe
strong spectral evolution from hard to soft. The spectral
evolution begins at T$_{\rm 0}$+285 s at $\gamma$-ray energies, while the
X-ray flux is
still at an approximately constant level, and begins 200 s later at X-ray
energies around the onset of the steep X-ray decay (Fig \ref{HR}). 

The spectral variability suggests a peak moving through the bandpass.
We find that the peak of the spectrum can be accurately measured by adding the
higher energy {\it Suzaku} WAM data to the {\it Swift} coverage. Assuming the Band function to model the spectrum, the
spectral peak started out at the high energy end of the BAT range
--- and was probably above 200 keV before the XRT or {\it Suzaku} observations
began --- and moved to lower energies to most probably lie within the XRT bandpass beyond
T$_{\rm 0}$+700 s (Fig. \ref{Epkband}). 
Indeed such strong spectral evolution may help explain some
of the steep X-ray decline. We also find an indication that the BAT spectral
lags may have decreased
during the prompt emission from the first main peak to $\sim$200 s later: such
lag evolution appears to be common among spectrally evolving GRBs (J. Cannizzo
et al. in preparation). The second, lower lag measurement was taken just after the
BAT hardness ratio had begun its
decrease and the spectral peak energy had fallen below 100 keV to be clearly
detectable within the BAT energy range.

While the spectral peak is moving we also measure a softening of the
spectrum at frequencies below the peak.
The spectral
index above the peak energy, $\beta$, is found to be 2.36 from the {\it
  Suzaku} data \citep{Morigami}. The spectrum is observed to be
harder at higher count rates, as has
been shown for several other GRBs (e.g. GRB\,051117a, Goad et al. 2007a;
GRB\,061121, Page et al. 2007). Superimposed on this general trend is a
hardening during the rise of individual flares, but this effect is not driving
the overall evolution (Fig. \ref{HRvsCR}).

Spectral evolution through the prompt phase has been noted previously, and is inconsistent with the idea that
the curvature effect alone -- the delayed arrival of emission from
progressively higher latitudes within the jet -- is driving the
emission during the steep decay phase \citep[e.g.][]{Liang}. If we try to force a curvature
effect model to explain the observed steep X-ray slope from the peak of the
last flare at T$_{\rm 0}$+757 s out to T$_{\rm 0}$+950 s, we would expect that phase to begin at T$_{\rm 0}$+632$^{+11}_{-12}$ s given the requirement of
the model that $\alpha=2+\beta$ ($F_{\nu} \propto t^{-\alpha} \nu^{-\beta}$
and adopting the average spectral energy index over that time interval of $\beta=1.06$, Fig. \ref{Tzero}). This would require a long initial emission period
of over 600 s. In the sample of {\it Swift} GRBs studied
by \cite{Liang} the typical injection of energy tends to occur at
the onset of the previous large peak.
The curvature effect clearly cannot explain the steepest part of the X-ray decay from
T$_{\rm 0}$+976--1200s (see last point in Fig. \ref{Tzero}, also seen in e.g. GRB\,051117a, Goad et
al. 2007a; GRB\,070110, Troja et al. 2007). However, the combination of the curvature
effect and the strong spectral evolution
we observe, which is a continuous process starting well before 600 s,
may be able to account for the steep X-ray decay.

\cite{Zhang2} studied the X-ray tails of the prompt emission
for a sample of {\it Swift} GRBs and found that only 25\% of those could be
fit with the curvature effect alone; those not well fit showed spectral evolution. From fits to 16 GRBs, they tested and subsequently disfavoured
two possible causes of the spectrally evolving X-ray tails, namely an
angle-dependent spectral index in structured jets and a superposition of the
curvature effect and a power law decay component.
The observed spectral softening could, they suggest, be caused by cooling of the
plasma where the cooling frequency decreases with time. This manifests itself as a cut-off power law shape with the
cut-off moving to lower energies with time, shown to be a good fit to
GRB\,060218 and proposed earlier for GRB\,980923
\citep{Giblin1}, and similar to our Band function results for GRB\,070616. In this
scenario the peak energy we track through the BAT band would be the
cooling frequency. The prompt emission SED at 770 s after the trigger,
including optical emission, is consistent with an absorbed broken power law
where the break shows the characteristics of a cooling break (though a single absorbed power law
provides an equally good fit). In the afterglow SED we most
probably do detect the
cooling break, hence we may have observed it moving to lower energies with
time. However, we also observe softening of the low energy power law slope in
addition to movement of a peak or cooling frequency.
Other models for the prompt emission
have been suggested - see \cite{Zhang1} for a review. 

\begin{figure}
\begin{center}
\includegraphics[width=6cm, angle=-90]{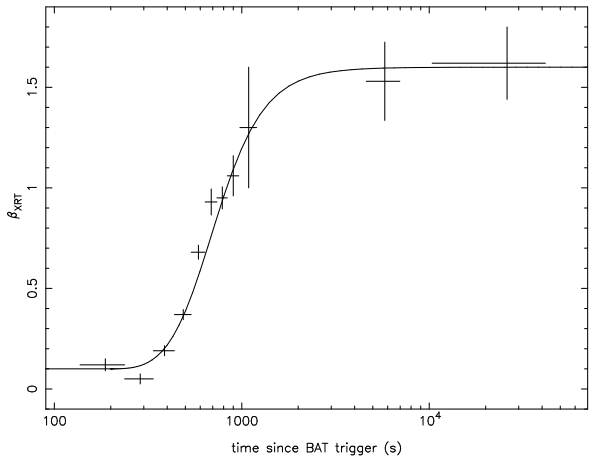}
\caption{Spectral evolution in the XRT band, plotted as $\beta$ vs. time for a
  direct comparison with Zhang et al. (2007),
where $\beta$ is the power law energy index of a single power law fit to the XRT spectra. We model this with the curvature model plus an underlying afterglow component to the strong spectral evolution (solid line). }
\label{curvatureplus}
\end{center}
\end{figure}
\begin{figure}
\begin{center}
\includegraphics[width=7cm, angle=-90]{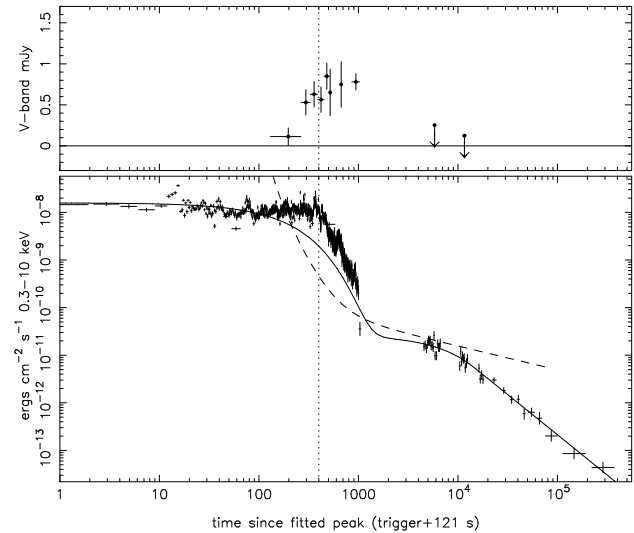}
\caption{Top panel: UVOT $V$ band flux light curve. Lower panel: BAT-XRT light curve 
  with two-component model fit
  following Willingale et al. (2007, solid line). This two-component model has proven a
  good fit to the majority of {\it Swift} high energy light curves, but is unable
  to fit all the prompt emission of GRB\,070616. The dashed line shows the flux profile
from the double power law model (equation 2) used to fit the $\beta$ profile
(Fig. \ref{curvatureplus}). This model also cannot fit the observed light curve.}
\label{dicks fit}
\end{center}
\end{figure}

We now investigate
origins for the apparent spectral evolution in the simple superposition of multiple components.
Fig. \ref{curvatureplus} shows the spectral evolution in the XRT band,
characterised by the fitted power law slope $\beta$ ($\beta = \Gamma -1$), displaying a
similar behaviour to many of the bursts studied by \cite{Zhang2} and resembling most closely that of GRB\,060510B which also had a lengthy prompt
emission duration of T$_{\rm 90}=275$ s with some flaring \citep{Barthelmy2}.
To approximate the shape of the spectral evolution of GRB\,070616 in the
X-ray band ($\beta$ vs. time) we fit a model with two power laws:
\begin{equation}
F_{E} = A_1 t^{-\alpha_1} E^{-\beta_1} + A_2 t^{-\alpha_2} E^{-\beta_2}
\end{equation}
where $t$ is time since a fitted T$_{\rm 0,fit}$ value and $E$ is the photon energy.
We fixed the spectral slopes to those of the typical measured prompt and afterglow
values, $\beta_1 = 0.1$ and $\beta_2 = 1.6$. We fixed $\alpha_2$ at the
probable afterglow temporal slope during the plateau
of 0.5 (see Section 3) while $\alpha_1$,
T$_{\rm 0,fit}$ and
the relative contributions of the prompt and afterglow components
were left to vary. 
This double power law
model, overlaid on the evolution of $\beta$ in Fig. \ref{curvatureplus}, appears to be a
reasonable description of the behaviour through the steep decay and afterglow
phases, with $\alpha_1=3.0$, T$_{\rm 0,fit}$ = T$_{\rm 0}$+200 s and the two components contributing
equally to the flux at T$_{\rm 0}$+800 s. The spectral shape evolves due to the
increasing contribution from the afterglow to the overall spectrum.
We then returned to the high energy light curve to
test the double power law model fitted above. The dashed line in the lower
panel of Fig. \ref{dicks fit} shows the double power law model applied to
the high energy light curve. Clearly, although the evolution of $\beta$ is well
modelled the flux is not. We can therefore rule out a
{\it spectrally invariant} double power law model.

It is possible that there is a further emission component, in addition to the high
latitude emission, contributing to the steep decay ending the prompt
emission, which becomes prominent between the prompt and early afterglow phases.
To examine this possibility 
we applied the modelling procedure of \cite{O'Brien} and \cite{Willingale} in which it was shown
that GRB early light curves can be well modelled by up to two emission components
each consisting of an exponential followed by a power law decay. The first
component models the prompt emission, adjusting the zero time, T$_{\rm 0}$, to provide the
best fit. For GRB\,070616 this moves zero close to the brightest peak in the
prompt light curve at 121 seconds after the trigger. The best-fit model
to the {\it Swift} BAT and XRT data is given in Table \ref{tab:lcfits2} and Fig.
\ref{dicks fit}.
\begin{table}
\caption{A two-component fit to the BAT-XRT light curve following Willingale
  et al. (2007). The trigger time has been rescaled to T$_{\rm 0}$+121 s. Quantities $\alpha_p$ and $T_p$ refer
to the prompt component power law decay slope and start of that decay
respectively, and $\alpha_a$ and $T_a$ refer
to the afterglow component, relative to the new zero time.}
\begin{center}
\renewcommand{\arraystretch}{1.5}
\begin{tabular}{ccl}
component& parameter & value\\ \hline \hline
1&$\alpha_p$& 7.37$^{+0.12}_{-0.27}$ \\
1&$T_p$    & 1400$^{+214}_{-156}$ s\\
2&$\alpha_a$& 1.69$^{+0.23}_{-0.20}$ \\
2&$T_a$     &14380$^{+5740}_{-4580}$ s \\\hline
\end{tabular}
\label{tab:lcfits2}
\end{center}
\end{table}

We find that the light curve of GRB\,070616 is
not well fitted by the usual two components consisting of an
exponential plus a power law decline. The first, prompt component can
fit the initial prompt data following the brightest peak, but then
declines until the second, afterglow component starts to
dominate. This final component can be fit with a power
law of slope $\alpha \sim 1.7$ from 4600 s onwards, typical of a post-plateau
GRB afterglow (Table \ref{tab:lcfits2}).
There remains an excess flux dominating from some 200--1000
seconds post-trigger. 
We speculate that this flux plus the spectral
evolution indicate an additional source of prompt emission which does
not dominate in the majority of GRBs. 

Interestingly, several, although not all,
of the GRBs with strong spectral evolution studied in \cite{Zhang2} are also
not well fitted by the two-component light curve model
\citep[e.g. GRBs 051227 and 060614, see][]{Willingale}. Others not
well fitted by this modelling procedure, such as GRB\,051117A \citep{Goad1} were
identified by \cite{Zhang2} as having evolution which
they attributed to flares. Such GRBs may instead be more similar to
GRB\,070616. GRB\,051117A, for example, shows a long, bright, slowly declining
prompt emission period with flaring followed by a sharp drop to a
plateau or afterglow phase. 
GRB\,070616 may be a case where
the flux remains fairly constant, with flaring, up to the sharp drop.
Among the {\it Swift} BAT bursts we notice of order 10 bursts with an initially flat
high energy light curve, most prominent in GRB\,070616, which when studied as
a sample may provide more evidence in support of either a long-duration
central engine or an additional component (N. Lyons et al. in preparation).

\subsection{The origin of the early optical emission}
The {\it Swift} era has provided a wealth of information on the early
$\gamma$-ray and X-ray emission. The prompt optical emission has been
harder to reveal as in most GRBs the flux level lies below
that accessible to fast-response, modest aperture telescopes. Even so,
for those GRBs which have been detected the relationship between the
optical and X-ray emission is complex. In a very small number of cases the
optical actually lies above the extrapolation of the higher-energy
emission \citep[e.g. GRB\,990123,][]{Akerlof}. The
optical can be dominated by a component related to the prompt high
energy emission \citep[e.g.][]{Vestrand1,Yost1}. More usually, the optical lies below such
an extrapolation, as shown for a sample of ROTSE-detected {\it Swift} GRBs by
Yost et al. (2007b, using BAT data only for the extrapolations). This pattern
of diverse behaviour strongly suggests a mixture of components can contribute to the prompt optical
emission, including the afterglow (due to the external shock) and
emission from the prompt fireball (which may include a reverse shock). In the latter case the optical
should connect, possibly with a spectral break, with the prompt
$\gamma$-ray -- X-ray spectrum.

Few optical datasets have sufficient broadband data to check for
consistency in spectral shape as well as overall flux level, limiting
our ability to constrain reddening effects as opposed to spectral
breaks. For GRB\,070616 the combination of long-duration prompt
emission plus early optical multi-colour data can provide some constraints on the
relationship between the optical emission and the high-energy prompt emission.
The $V$ band light curve shows a rise to a plateau-like phase from the first
observation beginning at T$_{\rm 0}$+250 s and extending to T$_{\rm 0}$+1000 s. The
optical decay does not follow the steeply decaying X- and
$\gamma$-ray prompt emission. A gradual rise would be typical of
the start of afterglow emission, seen in a number of bursts
\citep{Molinari}. We performed a power law fit to the $V$ band light curve
shown in Fig. \ref{all lightcurves} and measure $\alpha$~=~-($0.7\pm0.3$). This is
consistent with predictions for the afterglow rise in the fireball model of $\alpha=-0.5$ \citep[e.g.][]{zhangmeszaros}. 
Fits to the three prompt emission SEDs for GRB\,070616 all show that
extrapolation of the high energy spectrum to lower energies
overpredicts the optical flux, however this may be overcome with a relatively large
amount of intrisic extinction or a combination of extinction and a spectral
break, if the optical data are assumed to be prompt emission dominated.

We can compare this burst with other
GRBs for which well-time-resolved prompt optical data are available, and whose
light curves are not afterglow-dominated from early times.
For GRB\,050820A \cite{Vestrand2} find a
rising optical component plus a component correlated with the
$\gamma$-ray emission. In GRB\,050904 the optical data rose to a plateau
before mimicking a flare seen in the X-rays \citep{Boer}. Similarly in GRB\,051111 \cite{Yost1}
suggest a combination of components produce the prompt optical
emission with a significant contribution to the early optical flux from an
extension of the higher-energy spectrum. These are similar to the
case of GRB\,061121 \citep{Page} where the high energy prompt
emission consisted of a single large peak, which was mimicked in the
optical data, although an additional spectrally evolving optical
component is also seen. Extrapolation
of the high energy spectrum in GRB\,061121 overpredicts the early
optical emission, suggesting a spectral break. We note that the extrapolation
of the high energy spectrum to the optical band predicts a relatively higher
optical flux in GRB\,061121 than in GRB\,070616, due to its softer high energy
spectrum. 

For GRB\,070616 we conclude that strong optical emission associated with the
prompt phase and described by a simple extrapolation of the observed higher
energy spectrum cannot be present without invoking a further spectral break in
between optical and X-rays or extinction in the host at the highest end of the
currently observed values \citep[e.g.][]{Schady,Starling2}. 
We can likely rule out the $``$external-external" shock model \citep{Meszaros1}, which requires that
the broadband prompt emission is produced by one and the same mechanism and any
variability is caused by density enhancements in the external medium, given
the very different behaviour of the optical and X-ray light curves. 

\subsection{The afterglow}
After T$_{\rm 0}$+4000 seconds the X-ray afterglow
can be considered `typical' of what we know for GRB afterglows: the spectral
and temporal parameters lie within the range of the
$\sim$250 {\it Swift}-observed GRBs. We test these
parameters against the synchrotron model, using the
closure relations for the case of a cooling break in between the optical and
X-ray bands, $\nu_c < \nu_X$ and $\alpha = (3\Gamma-4)/2$, and find that the observed temporal and spectral X-ray slopes
are just consistent at the 90\% level and therefore follow the expectations for synchrotron
emission (in the slow cooling regime). This result also indicates that the afterglow has not reached the
`jet' phase,
in agreement with the lack of evidence for a jet break during the {\it Swift}
observations. We do not find good agreement if the cooling frequency were to
lie above the X-ray band, $\nu_c > \nu_X$, at this time, consistent with the requirement of
a broken power law to best describe the SED of the
late-time afterglow spectrum plus optical upper limits. Therefore we conclude that the
observed spectral break is due to the cooling frequency. The cooling frequency
is likely to lie at or above the X-ray frequencies at earlier times ($<$4000
s), depending on the circumburst medium density structure, but we cannot
confirm this since the initial X-ray emission is prompt-dominated.

We find a 90\% confidence range for
the injected electron energy index, $p$, of 2.68--3.46, as determined from the
measured X-ray spectral index where $\Gamma = (p+2)/2$. This is higher
than the commonly quoted `universal' value of $p=2.2$ derived from some
numerical simulations of particle acceleration in relativistic shocks
\citep[e.g.][]{Achterberg}, but other numerical simulations \citep[e.g.][]{Baring} and
recently published observational studies of samples of afterglows
\citep{Shen,Starling3} find a range of allowed values for
$p$ which encompass the values we report for GRB\,070616.

\section{Conclusions}
GRB\,070616 has afforded us the rare opportunity to track the detailed
spectral and temporal evolution from $\gamma$-ray to optical wavelengths throughout
much of the prompt emission phase, owing to {\it Swift's} fast slew
capability, the broadband coverage provided by the combination of {\it Swift}
with {\it Suzaku} WAM, and
the long prompt emission duration.
The high energy light curve remains generally flat for several hundred seconds
before going into a steep decline.

Spectral evolution from hard to soft is clearly taking place through the
prompt emission (flat and decaying phases) of GRB\,070616, beginning at 285 s
out to at least 1000 s. The high energy spectrum from 0.3--800 keV is well
modelled by a Band function. We track the spectral peak energy moving from 135
keV down to $\sim$4 keV and measure a softening of the low energy spectral slope from $\Gamma$$\sim$1.1--2.3.
The curvature effect, whilst contributing to the spectral shape, is not
dominating during the steep decay at $\sim$1000 s. 

The presence of an additional component, perhaps not present in most GRBs, is favoured by the inability of a two-component fit to
model the light curve. This unusual light curve
shape of an initial constant flux followed by a sharp drop, plus the strong spectral evolution, is most prominent in GRB\,070616, but
we note a similar structure can be found in a small number of other GRBs which
are also not dominated by the curvature
effect and not well fit by the two-component model.

We find that the optical data during the prompt phase are not consistent with a simple extrapolation of the high
energy spectrum, requiring either significant
intrinsic extinction or some extinction
plus a
break in the spectrum between the X-ray and
optical bands.

The afterglow is consistent with the synchrotron
model, with the cooling break lying in between optical and X-ray bands at a
few thousand seconds after the trigger.

The prompt emission of GRB\,070616 comprises a component well fitted with a
Band function and a possible further component. 
It is clear that both broadband coverage and good time resolution are crucial to
pinning down the origins of the complex prompt emission in GRBs.

\begin{figure*}
\begin{center}
\includegraphics[width=18cm, angle=0]{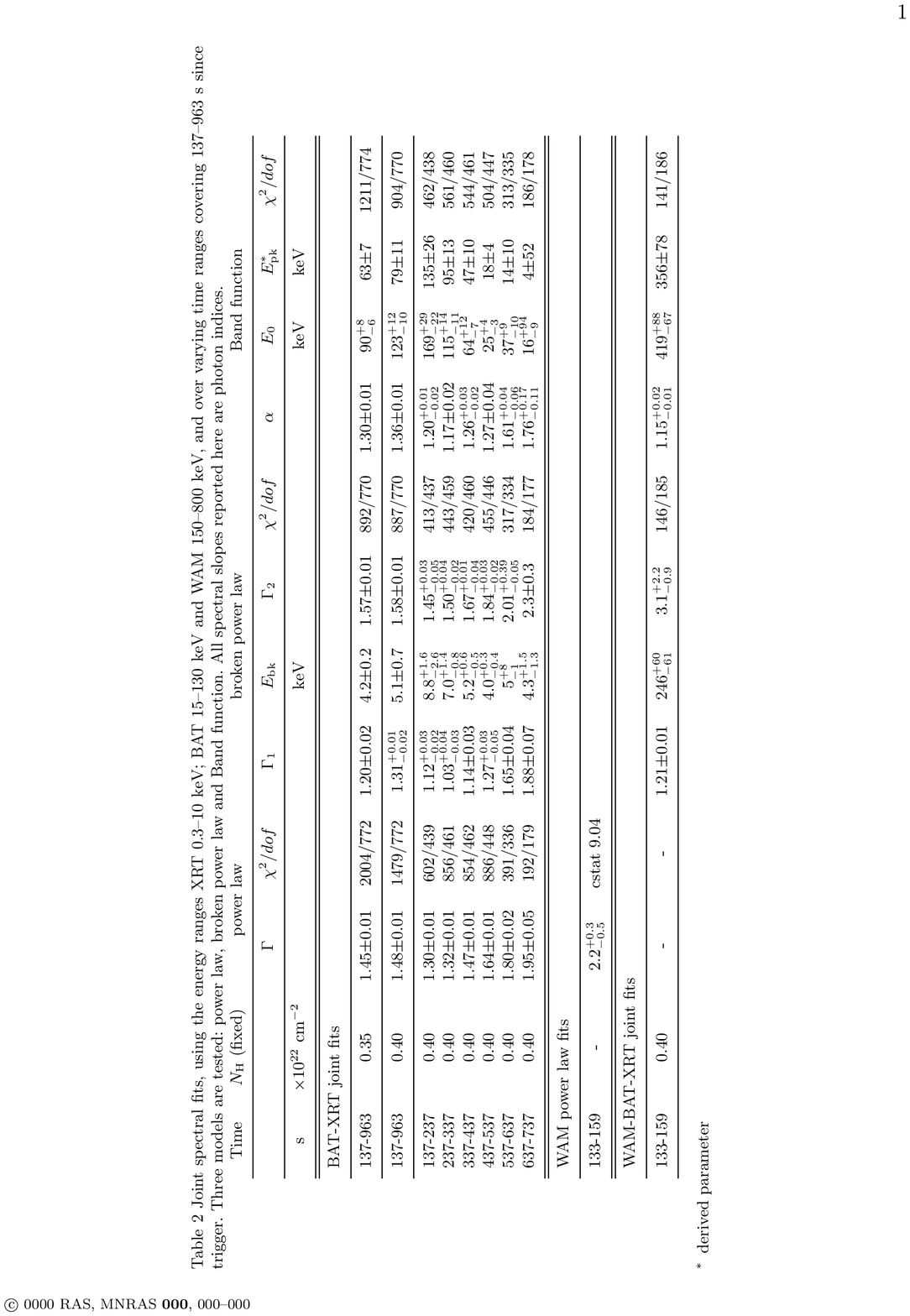}
\label{replacement}
\end{center}
\end{figure*}

\section{Acknowledgments}
We thank E. Rol,
O. Godet, P.A. Evans, G. Sato, S.T. Holland, F.E. Marshall,
S.R. Oates and A. de Ugarte Postigo for useful
discussions. We thank all the {\it Suzaku} WAM team for their contributions,
in particular M.S. Tashiro, K. Yamaoka,
K. Morigami and M. Ohno, and we acknowledge the whole {\it Swift} team for
their many contributions. We thank B. Gendre for useful comments on the
manuscript. RLCS, KLP and JPO acknowledge financial support from STFC.

\bsp

\label{lastpage}


\begin{thebibliography}{References}
\bibitem[\protect\citeauthoryear{Achterberg et al.}{2001}]{Achterberg}
Achterberg A., Gallant Y.A., Kirk J.G., Guthmann A.W., 2001, MNRAS, 328, 393 
\bibitem[\protect\citeauthoryear{Akerlof et al.}{1999}]{Akerlof}
Akerlof C. et al., 1999, Nature, 398, 400 
\bibitem[\protect\citeauthoryear{Alard \& Lupton}{1998}]{Alard}
Alard C., Lupton R.H., 1998, ApJ, 503, 325
\bibitem[\protect\citeauthoryear{Arnaud}{1996}]{Arnaud}
Arnaud K., 1996, ASPC, 101, 17 
\bibitem[\protect\citeauthoryear{Band et al.}{1993}]{Band}
Band D. et al., 1993, ApJ, 413, 281 
\bibitem[\protect\citeauthoryear{Baring}{2004}]{Baring}
Baring M.G., 2004, Nuclear Physics B Proceedings Supplements, 136, 198 
\bibitem[\protect\citeauthoryear{Barthelmy et al.}{2005}]{Barthelmy}
Barthelmy S. et al., 2005, Space Sci. Rev., 120, 143 
\bibitem[\protect\citeauthoryear{Barthelmy et al.}{2006}]{Barthelmy2}
Barthelmy S. et al., 2006, GCN Circulars, 5107
\bibitem[\protect\citeauthoryear{Bo\"er et al.}{2006}]{Boer}
Bo\"er M., Atteia J.L., Damerdji Y., Gendre B., Klotz A., Stratta G., 2006,
ApJ, 638, L71
\bibitem[\protect\citeauthoryear{Burrows et al.}{2005a}]{Burrows2}
Burrows D.N. et al., 2005a, Science, 309, 1833
\bibitem[\protect\citeauthoryear{Burrows et al.}{2005b}]{Burrows1}
Burrows D.N. et al., 2005b, Space Sci. Rev., 120, 165 
\bibitem[\protect\citeauthoryear{Campana et al.}{2006a}]{Campana1}
Campana S. et al., 2006a, A \& A, 449, 61  
\bibitem[\protect\citeauthoryear{Campana et al.}{2006b}]{Campana2}
Campana S. et al., 2006b, Nature, 442, 1008 
\bibitem[\protect\citeauthoryear{Chincarini et al.}{2007}]{Chincarini}
Chincarini G. et al., 2007, ApJ in press, preprint (arXiv:astro-ph/0702371) 
\bibitem[\protect\citeauthoryear{Falcone et al.}{2007}]{Falcone}
Falcone A.D. et al., 2007, ApJ submitted, preprint (arXiv:0706.1564)
\bibitem[\protect\citeauthoryear{Gehrels et al.}{2004}]{Gehrels1}
Gehrels N. et al., 2004, ApJ, 611, 1005 
\bibitem[\protect\citeauthoryear{Gehrels et al.}{2005}]{Gehrels2}
Gehrels N. et al., 2005, Nature, 437, 851 
\bibitem[\protect\citeauthoryear{Giblin et al.}{2002}]{Giblin}
Giblin T.W., Connaughton V., van Paradijs J., Preece R.D., Briggs M.S., Kouveliotou C., Wijers R.A.M.J., Fishman G.J., 2002, ApJ, 570, 573 
\bibitem[\protect\citeauthoryear{Giblin et al.}{1999}]{Giblin1}
Giblin T.W., van Paradijs J., Kouveliotou C., Connaughton V., Wijers R.A.M.J.,
Briggs M.S., Preece R.D., Fishman G.J., 1999, ApJ, 524, L47
\bibitem[\protect\citeauthoryear{Goad et al.}{2007a}]{Goad1}
Goad M.R. et al., 2007a, A \& A, 468, 103 
\bibitem[\protect\citeauthoryear{Goad et al.}{2007b}]{Goad2}
Goad M.R. et al., 2007b, A \& A in press, preprint (arXiv:0708.0986)
\bibitem[\protect\citeauthoryear{Houck \& Denicola}{2000}]{Houck}
Houck J.C., Denicola L.A., 2000, ASPC, 216, 591 
\bibitem[\protect\citeauthoryear{in 't Zand et al.}{2000}]{Zand}
in 't Zand J.J.M. et al., 2004, Proceedings of the Third Rome Workshop on Gamma-Ray Bursts in the Afterglow Era, ASP
Conference Series, 312, 18, Eds. M. Feroci, F. Frontera, N. Masetti, L. Piro. 
\bibitem[\protect\citeauthoryear{Kalberla et al.}{2005}]{Kalberla}
Kalberla P.M.W., Burton W.B., Hartmann D., Arnal E.M., Bajaja E., Morras R., Poppel W.G.L., 2005, A \& A, 440, 775 
\bibitem[\protect\citeauthoryear{Kann \& Wilson}{2007}]{Kann}
Kann D.A., Wilson A.C., 2007, GCN Circulars, 6629 
\bibitem[\protect\citeauthoryear{Krimm et al.}{2007}]{Krimm}
Krimm H. et al., 2007, GCN Circulars, 6058 
\bibitem[\protect\citeauthoryear{Kumar et al.}{2007}]{Kumar1}
Kumar P. et al., 2007, MNRAS, 376, L57 
\bibitem[\protect\citeauthoryear{Kumar \& Panaitescu}{2000}]{Kumar2}
Kumar P., Panaitescu A., 2000, ApJ, 541, L51
\bibitem[\protect\citeauthoryear{Liang et al.}{2006}]{Liang}
Liang E.W. et al., 2006, ApJ, 646, 351
\bibitem[\protect\citeauthoryear{M\'esz\'aros \& Rees}{1993}]{Meszaros1}
M\'esz\'aros P., Rees, M.J., 1993, ApJ, 405, 278
\bibitem[\protect\citeauthoryear{M\'esz\'aros \& Rees}{1997}]{Meszaros2}
M\'esz\'aros P., Rees, M.J., 1997, ApJ, 476, 232 
\bibitem[\protect\citeauthoryear{M\'esz\'aros, Rees \& Papathanassiou}{1994}]{Meszaros3}
M\'esz\'aros P., Rees, M.J., Papathanassiou, H., 1994, ApJ, 432, 181 
\bibitem[\protect\citeauthoryear{Mitsuda et al.}{2007}]{Mitsuda}
Mitsuda K. et al. 2007, PASJ, 59, S1 
\bibitem[\protect\citeauthoryear{Molinari et al.}{2007}]{Molinari}
Molinari E. et al., 2007, A \& A, 469, L13  
\bibitem[\protect\citeauthoryear{Morigami et al.}{2007}]{Morigami}
Morigami K. et al., 2007, GCN Circulars, 6578 
\bibitem[\protect\citeauthoryear{Nicastro et al.}{2004}]{Nicastro}
Nicastro L. et al., 2004, A\&A, 427, 445
\bibitem[\protect\citeauthoryear{Norris et al.}{1996}]{Norris}
Norris J.P., Nemiroff R.J., Bonnell J.T., Scargle J.D., Kouveliotou C., Paciesas W.S., Meegan C.A., Fishman G.J., 1996, ApJ, 459, 393
\bibitem[\protect\citeauthoryear{Nousek et al.}{2006}]{Nousek}
Nousek J. et al., 2006, ApJ, 642, 389  
\bibitem[\protect\citeauthoryear{O'Brien et al.}{2006}]{O'Brien}
O'Brien P.T. et al., 2006, ApJ, 647, 1213 
\bibitem[\protect\citeauthoryear{Paciesas et al.}{1999}]{Paciesas}
Paciesas W.S. et al., 1999, ApJS, 122, 465 
\bibitem[\protect\citeauthoryear{Page et al.}{2007}]{Page}
Page K.L. et al., 2007, ApJ, 663, 1125 
\bibitem[\protect\citeauthoryear{Palmer et al.}{2006}]{Palmer}
Palmer D., Sato G., Barthelmy S.D., Sakamoto T., Markwardt C., Gehrels N., 2006, GCN Circulars, 5662  
\bibitem[\protect\citeauthoryear{Price et al.}{2002}]{Price}
Price P.A. et al., 2002, ApJ, 573, 85  
\bibitem[\protect\citeauthoryear{Rees \& M\'esz\'aros}{1994}]{Rees}
Rees M.J., M\'esz\'aros P., 1994, ApJ, 430, L93 
\bibitem[\protect\citeauthoryear{Roming et al.}{2005}]{Roming}
Roming P.W.A. et al., 2005, Space Sci. Rev., 120, 95 
\bibitem[\protect\citeauthoryear{Sakamoto et al.}{2007}]{Sakamoto}
Sakamoto T. et al., 2007, ApJ in press, preprint (arXiv:0707.4626) 
\bibitem[\protect\citeauthoryear{Schady et al.}{2007}]{Schady}
Schady P. et al., 2007, MNRAS, 377, 273
\bibitem[\protect\citeauthoryear{Schlegel, Finkbeiner \& Davis}{1998}]{Schlegel}
Schlegel D.J., Finkbeiner D.P., Davis M., 1998, ApJ, 500, 525 
\bibitem[\protect\citeauthoryear{Shen, Kumar \& Robinson}{2006}]{Shen}
Shen R., Kumar P., Robinson E.L., 2006, MNRAS, 371, 1441 
\bibitem[\protect\citeauthoryear{Starling et al.}{2007a}]{Starling1}
Starling R.L.C. et al., 2007a, GCN Circulars, 6542 
\bibitem[\protect\citeauthoryear{Starling et al.}{2007b}]{Starling2}
Starling R.L.C., Wijers R.A.M.J., Wiersema K., Rol E., Curran P.A., Kouveliotou C., van der Horst A.J., Heemskerk M.H.M., 2007b, ApJ, 661, 787 
\bibitem[\protect\citeauthoryear{Starling et al.}{2007c}]{Starling3}
Starling R.L.C., van der Horst A.J., Rol E., Wijers R.A.M.J., Kouveliotou
C., Wiersema K., Curran P.A., Weltevrede P., 2007c, ApJ in press, preprint (arXiv:0704.3718)   
\bibitem[\protect\citeauthoryear{Tagliaferri et al.}{2005}]{Tagliaferri}
Tagliaferri G. et al., 2005, Nature, 436, 985 
\bibitem[\protect\citeauthoryear{Takahashi et al.}{2007}]{Takahashi}
Takahashi T. et al., 2007, PASJ, 59, 35
\bibitem[\protect\citeauthoryear{Troja et al.}{2007}]{Troja}
Troja E. et al., 2007, ApJ, 665, 599
\bibitem[\protect\citeauthoryear{Usov}{1994}]{Usov}
Usov V.V., 1994, MNRAS, 267, 1035 
\bibitem[\protect\citeauthoryear{Vestrand et al.}{2005}]{Vestrand1}
Vestrand W.T. et al., 2005, Nature, 435, 178 
\bibitem[\protect\citeauthoryear{Vestrand et al.}{2006}]{Vestrand2}
Vestrand W.T. et al., 2006, Nature, 442, 172 
\bibitem[\protect\citeauthoryear{Willingale et al.}{2007}]{Willingale}
Willingale R. et al., 2007, ApJ, 662, 1093 
\bibitem[\protect\citeauthoryear{Yamaoka et al.}{2005}]{Yamaoka}
Yamaoka K. et al., 2005, IEEE, TNS, Vol. 52, No. 6, 2765
\bibitem[\protect\citeauthoryear{Yost et al.}{2007a}]{Yost1}
Yost S.A. et al., 2007a, ApJ, 657, 925 
\bibitem[\protect\citeauthoryear{Yost et al.}{2007b}]{Yost2}
Yost S.A. et al., 2007b, ApJ, 669, 1107
\bibitem[\protect\citeauthoryear{Zhang}{2007}]{Zhang1}
Zhang B., 2007, Chin. J. Astrophys, 7, 1 
\bibitem[\protect\citeauthoryear{Zhang et al.}{2007}]{Zhang2}
Zhang B.-B., Liang E.-W., Zhang B., 2007, ApJ, 666, 1002
\bibitem[\protect\citeauthoryear{Zhang \& M\'esz\'aros}{2004}]{zhangmeszaros}
Zhang B., M\'esz\'aros P., 2004, Int.J.Mod.Phys., A19, 2385
\end{thebibliography}
\end{document}